\documentclass[runningheads]{llncs}
\usepackage{graphicx, tabularx}
\newcolumntype{Y}{>{\centering\arraybackslash}X}
\usepackage[urlcolor=blue]{hyperref}

\usepackage{cite}

\usepackage{xcolor}
\usepackage{multirow}

\usepackage{amsmath}

\begin{document}
\title{Tempera: Spatial Transformer Feature Pyramid Network for Cardiac MRI Segmentation}
%
\titlerunning{Tempera Segmentation Network}
%
\author{Christoforos Galazis\inst{1, 2}* \and Huiyi Wu\inst{2} \and Zhuoyu Li\inst{3} \and Camille Petri\inst{2}  \and Anil A. Bharath\inst{4} \and Marta Varela\inst{2}}
%
\authorrunning{C. Galazis et al.}
%
\institute{
Department of Computing, Imperial College London, UK \and
National Heart \& Lung Institute, Imperial College London, UK \and
Department of Metabolism,Digestion \&Reproduction, Imperial College London, UK \and
Department of Bioengineering, Imperial College London, UK \\ 
\email{*c.galazis20@imperial.ac.uk}}

%
\maketitle              
\begin{abstract}
Assessing the structure and function of the right ventricle (RV) is important in the diagnosis of several cardiac pathologies. However, it remains more challenging to segment the RV than the left ventricle (LV). In this paper, we focus on segmenting the RV in both short (SA) and long-axis (LA) cardiac MR images simultaneously. For this task, we propose a new multi-input/output architecture, hybrid 2D/3D geometric spatial TransformEr Multi-Pass fEature pyRAmid (Tempera). Our feature pyramid extends current designs by allowing not only a multi-scale feature output but multi-scale SA and LA input images as well. Tempera transfers learned features between SA and LA images via layer weight sharing and incorporates a geometric target transformer to map the predicted SA segmentation to LA space. Our model achieves an average Dice score of 0.836 and 0.798 for the SA and LA, respectively, and 26.31mm and 31.19mm Hausdorff distances. This opens up the potential for the incorporation of RV segmentation models into clinical workflows.

\end{abstract}

\setlength\belowcaptionskip{-3.5ex}

\section{Introduction}
\label{label:introduction}
Cardiac Magnetic Resonance Imaging (MRI) is the most widely used imaging technique to quantify the structure and function of the heart \cite{cmr}. As such, it can be used to assess the right ventricle (RV) and contribute to the diagnosis and monitoring of cardiac pathologies such as coronary heart disease, pulmonary hypertension, dysplasia and cardiomyopathies \cite{Caudron2011DiagnosticAA, right_vent_assessment}. 

Despite the RV's importance, the left ventricle (LV) has traditionally been analysed in greater detail due to its pivotal role in a wider range of pathologies \cite{PETITJEAN2015187}. The RV is more challenging to accurately segment for both clinicians and (semi-)automated algorithms \cite{Bonnemains2012, PETITJEAN2015187}. This is due to the RV's more complex crescent shape, thinner ventricular wall and heavier presence of wall trabeculations compared to the LV \cite{rv_lv, PETITJEAN2015187}.

In this paper, we focus on RV automatic segmentation as part of the "\textit{Multi-Disease, Multi-View \& Multi-Center Right Ventricular Segmentation in Cardiac MRI}" (M\&Ms2) challenge \cite{carlos_martin_isla_2021_4573984}. We propose a novel hybrid 2D/3D deep neural network that takes both short-axis (SA) and long-axis (LA) images as inputs. It includes a novel multi-input/output feature pyramid that facilitates weight sharing across SA in-plane slices and between the SA and LA views. Additionally, we include a geometric target spatial transformer that utilizes the known spatial relationship between the different cardiac views.

\section{Methods}
\label{label:methods}
\textit{\textbf{Data}}
The end-diastolic (ED) and end-systolic (ES) cardiac MRI image phases used for the experiments come from a 360-subject bSSFP CINE MRI dataset publicly available through the M\&Ms-2 challenge \cite{carlos_martin_isla_2021_4573984}. The dataset is obtained across different sites, different vendors and has pathologies in the test set not included in the training. The available ground truth has manual segmentations of the LV and RV blood pool and the LV myocardium. From this set, 160 cases are dedicated for training. The remaining 40 and 160 cases are the development and test sets used for their respective phases in the competition. We further split the 160-case training set to 150 for model training and 10 for validation. 

\textit{\textbf{Preprocessing}}
\label{label:methods:preprocessing}
As the images were acquired across different centers and vendors, they need to be standardized before being passed to the model. The resolution of both SA and LA images was resampled to 1.25 $\times$ 1.25 mm in-plane using b-spline interpolation. The through-plane spacing was left unchanged at 10 mm. We first automatically identify a region of interest (ROI) containing the RV and LV by applying Canny edge detection on the ED and ES image difference, then using the circular Hough transform to identify the heart. This allows us to identify the regions in the image that have the largest movement across the cardiac cycle, which we assume to be the heart. The image size was standardized to 192 $\times$ 192 $\times$ 17 through center ROI cropping. Finally, we standardize the cropped image such that it has a mean of zero and unit variance.

Additionally, we identify the affine transformation between the SA and LA images, which will subsequently be used in the segmentation network. The LA is treated as a 3D image (by adding a depth axis of size 1), thus allowing for a 3D/3D registration to take place. We pre-align the images based on the available file metadata of the images. For the registration, we follow a coarse-to-fine blurring approach to initially align with global features before finer ones. The registration is optimized to maximize the mutual information score. This is done using the SimpleITK package.

\begin{figure}[ht!]
\centering
  \includegraphics[width=0.9\textwidth]{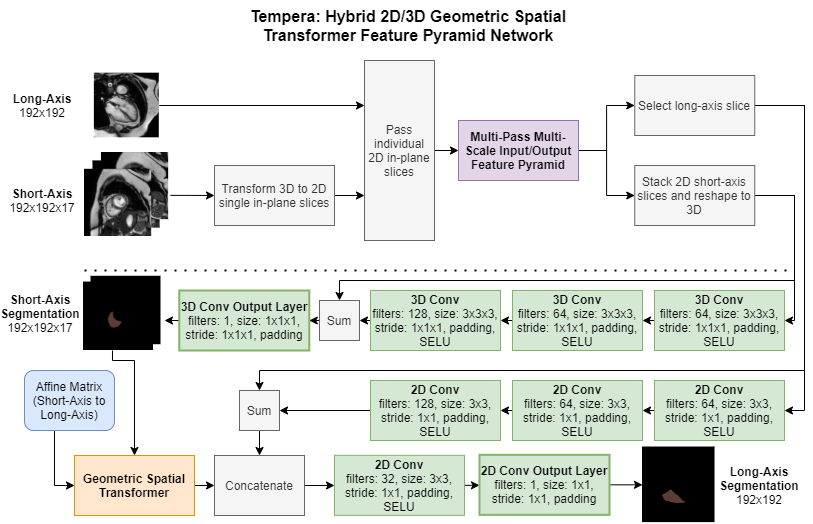}
  \caption{An overview of the architecture of the hybrid 2D/3D Tempera network. It receives the individual 2D in-plane slices for both the SA and LA as inputs to the feature pyramid component. Then, the short-axis slices are used to reconstruct the 3D representation and each image view is passed to their respective 2D/3D convolutional branch. Finally, the predicted SA segmentation is transformed to the LA space to improve its prediction.}
  \label{fig:network_architecture}
\end{figure}

\textit{\textbf{Model:Architecture}}
\label{label:methods:model}
To simultaneously segment the SA and LA images, we propose a new hybrid 2D/3D geometric spatial TransformEr Multi-Pass fEature pyRAmid (Tempera) network, as shown in detail in \textit{Figure \ref{fig:network_architecture}}. Our architecture consists of 2D SA/LA hard-weight sharing layers, independent 3D (SA) and 2D (LA) branches and finally a geometric spatial transformer.

\begin{figure}[ht!]
\centering
  \includegraphics[width=0.77\textwidth]{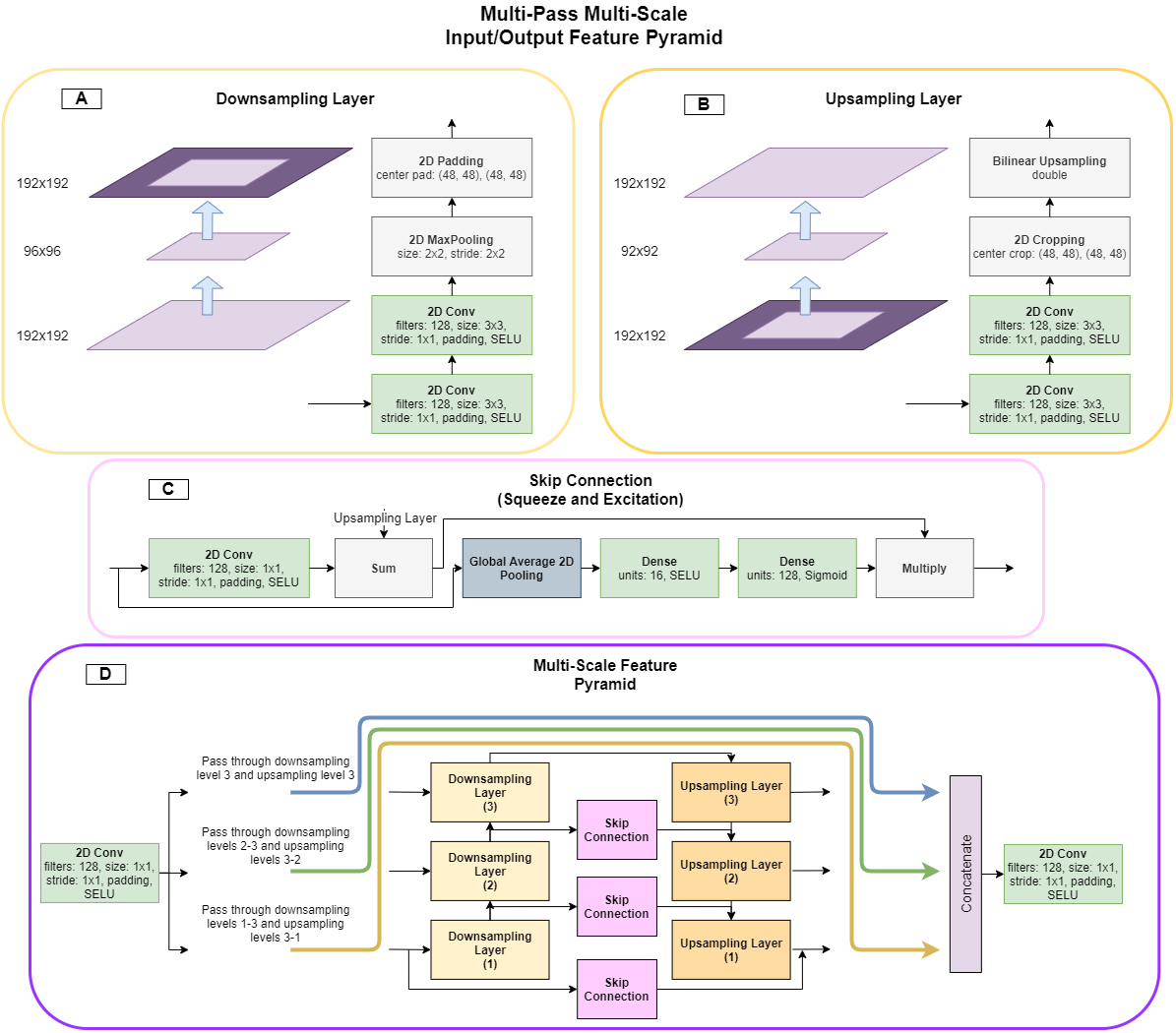}
  \caption{Architecture of our proposed Multi-Pass and multi-scale input/output Feature Pyramid (MPFP). The downsampling component in block \textit{A} consists of two 2D convolutional layers, followed by a pooling and padding layer. Similarly, the upsampling layer, block \textit{B}, has two 2D convolutional layers, a cropping operation and finally bilinear upsampling. The skip connection in \textit{C} contains a 2D convolutional layer followed by a summing operation and a squeeze-and-excitation component \cite{squeeze_and_excitation} to scale features. Finally, \textit{D} illustrates how the components interact to form the MPFP.  2D convolutional layers are used to generate consistent feature sizes for the inputs and also to merge the outputs. After every layer we use a Scaled Exponential Linear Unit (SELU) \cite{selu} activation function, with exception the last layer of the skip connection uses a sigmoid. The input data to the pyramid is passed $n$ times and generates $n$ outputs, where $n=3$ is the number of pyramid levels. Each subsequent pass goes through one less downsampling and upsampling layer.}
  \label{fig:pyramid_architecture}
\end{figure}

\begin{figure}[ht!]
\vspace{3mm}
\centering
  \includegraphics[width=0.5\textwidth]{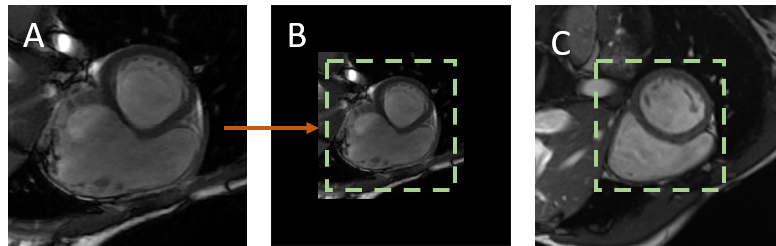}
  \caption{Illustrative examples of how the feature pyramid may allow a hypothetical feature transfer to an unknown pathology for the network (dilated RV). Image \textit{A} depicts SA basal slice of a patient with dilated RV. Image \textit{B} is the downsampled version of image \textit{A} by the MPFP in the first block. Finally, image \textit{C} shows the SA basal slice of a healthy subject. The RV in images \textit{B} and \textit{C} are roughly of equal size. This allows feature transfer to untrained dilated RV cases.}
  \label{fig:features_from_pyramid}
\end{figure}

For the shared layers, we extend the Multi-scale Feature Pyramid network \cite{feature_pyramid} to an architecture that we call Multi-Pass Feature Pyramid (MPFP). In the MPFP, the weight sharing, and thus internal (3x3) feature transfer, is achieved across the SA and LA in-plane slices and between the SA slices themselves. We are thus able to share features across all slices that relate to the RV shape, texture and contrast relative to the surroundings. The (partial) scale-invariant input/output of MPFP is designed to help Tempera generalize to unseen pathologies, such as dilated RV. An example is shown in \textit{Figure \ref{fig:features_from_pyramid}} in which we base our assumption. The specific implementation details of our pyramid can be viewed in \textit{Figure \ref{fig:pyramid_architecture}}.

\begin{figure}[ht!]
\centering
  \includegraphics[width=0.56\textwidth]{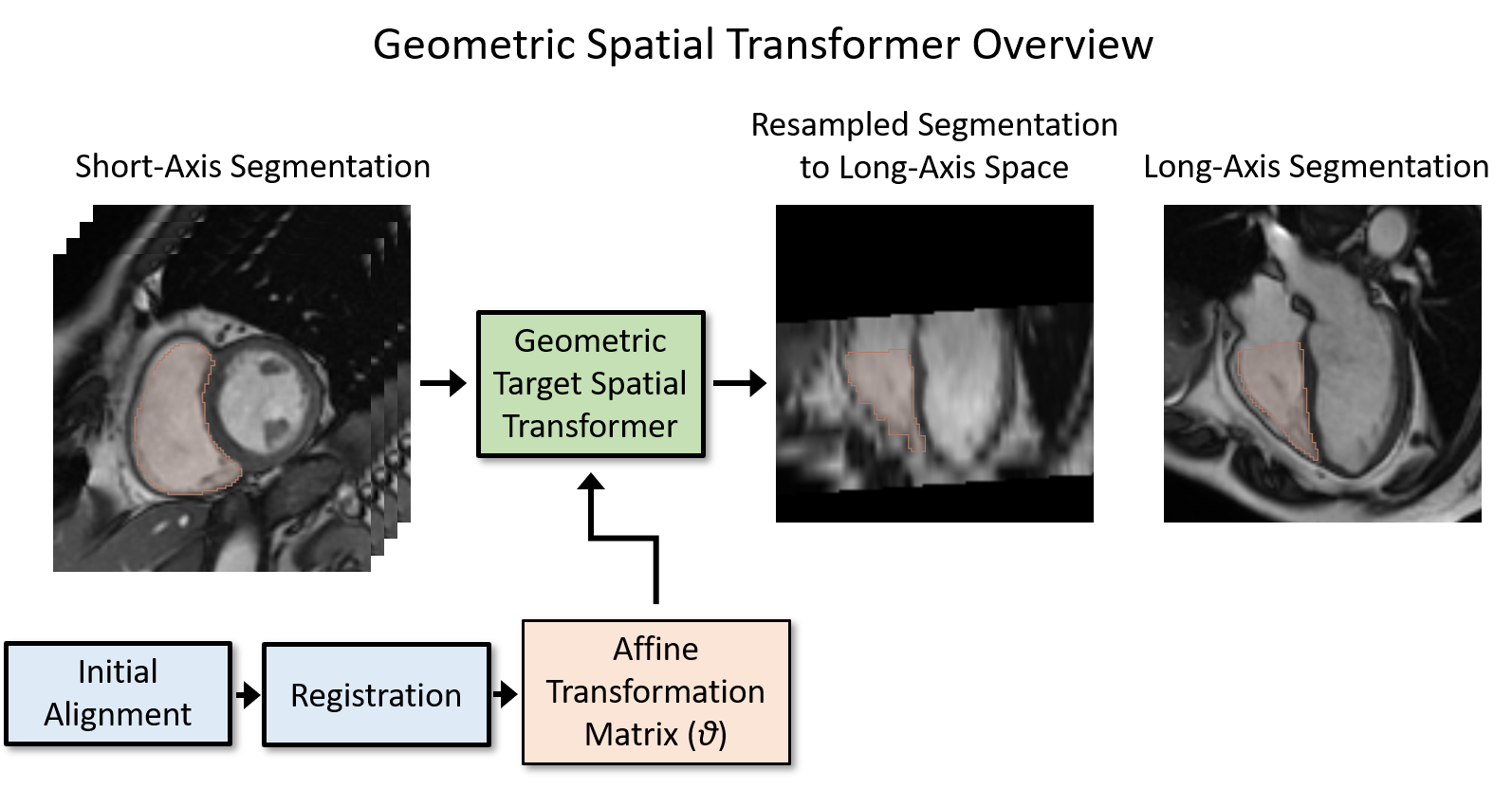}
  \caption{The Geometric target Spatial Transformer (GST) affine transforms and resamples using linear interpolation the predicted segmentation from the SA to LA space, given an affine matrix as input. Their spatial relationship is stored in the image's metadata, in which we further correct with an affine registration algorithm.}
  \label{fig:spatial_transformer}
\end{figure}

Our second component is the Geometric target Spatial Transformer (GST), depicted in \textit{Figure \ref{fig:spatial_transformer}}. This is a differentiable non-trainable component, that we have built upon the spatial transformer from \cite{voxelmorph}. It applies transformations between different domain and target spaces, such as 3D to 2D and vice versa. The GST takes as input the pre-computed affine matrix to perform relevant coordinate transformations from the SA to LA. Specifically, we utilise it on the predicted SA segmentation to localize the RV in the LA.

\textit{\textbf{Model:Optimizer}}
We use the Adam optimizer \cite{kingma2017adam} with: exponential decay rates for first moment estimates $\beta_1=0.9$ , second moment estimates $\beta_2=0.999$ and $\epsilon=10^{-8}$ to avoid division with 0. Additionally, we use an empirically-chosen initial learning rate of $5 \times 10^{-4}$, accompanied by a learning rate decay of 0.9 every 30 epochs.

\textit{\textbf{Model:Loss Function}}
We minimized the combined \cite{anatomynet} equal weighted Dice and focal loss (multiplicative weighted cross-entropy) \cite{lin2018focal} functions. We set the focal loss class weights $\alpha=0.25$ and relaxation term $\gamma=2.0$. Through our experimentation, we found that including the Dice loss improves the learning during initial epochs. The focal loss contributes during the later stages by dealing with class imbalance, in which background pixels outweigh foreground ones. Additionally, it helps with difficult to classify areas regardless if they are foreground or background. We set a $\lambda$ of 75 for the SA loss and 1 for the LA. The equation is formed as bellow, where $p$ is the probability of the ground truth label:

\setlength{\belowdisplayskip}{2pt} \setlength{\belowdisplayshortskip}{2pt}
\setlength{\abovedisplayskip}{1pt} \setlength{\abovedisplayshortskip}{1pt}

\begin{equation}
    L_{Dice(p_{t})} = 1 - \frac{2 \times true\_positive_{p_{t}}}{2 \times true\_positive_{p_{t}} + false\_positive_{p_{t}} + false\_negative_{p_{t}}}
\end{equation}

\begin{equation}
    L_{Focal}(p_{t}) = -a_{t} \times (1-p_{t})^{\gamma} \times log(p_{t})
\end{equation}

\begin{equation}
    \begin{split}
        L_{Total}(p_{SAt}, p_{LAt}) =& \lambda _{SA} \times (L_{Dice}(p_{SAt}) + L_{Focal}(p_{SAt}) + \\& \lambda_{LA} \times L_{Dice}(p_{LAt} + L_{Focal}(p_{LAt})
    \end{split}
\end{equation}

\textit{\textbf{Model:Hyperparameters}}
We use LeCun weight initialization \cite{lecun, selu} from a normal distribution and set the biases to zero. This is done to maintain the self-normalizing property when using Scaled Exponential Linear Units (SELUs) \cite{selu}. Also, we set the batch size to 1 due to memory limits and train for 300 epochs, which took nearly 16 hours to complete. The source code is available at \url{https://github.com/cgalaz01/mnms2\_challenge}.

\textit{\textbf{Model:Data Augmentation}}
To prevent the model from overfitting and to expand the dataset, we perform data augmentations. These include: in-plane rotations, in-plane anisotropic scalings, blurring, Gaussian noise addition, mean intensity shift and removing square segments of the image (in-painting). We use randomly selected parameters at each data iteration for each epoch.

\textit{\textbf{Model:Data Postprocessing}}
We found that our model will occasionally identify surrounding tissues as the RV. To circumvent this, we added a post-processing step, in which only the largest connected region will be considered as a valid prediction. Additionally, we apply a median filter to correct some of the interpolation errors caused by transforming back to the domain space using nearest neighbor interpolation.

\section{Results}
Tempera is able to accurately segment both the SA and LA views. On the test set we obtain 0.836 $\pm$ 0.23 Dice score and 26.31 $\pm$ 52.68 mm Hausdorff distance for the SA. The LA has a lower performance with 0.798 $\pm$ 0.28 Dice score and 31.19 $\pm$ 31.19 mm Hausdorff distance. From the predicted tests cases, there were a total of 11 SA and 20 LA that most likely failed during the ROI detection process, thus returning empty segmentations. Full details of the results are shown in \textit{Table \ref{tab:results}}.

\begin{figure}[ht!]
\centering
  \includegraphics[width=0.9\textwidth]{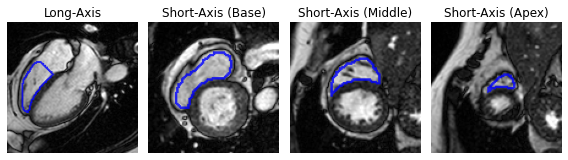}
  \caption{A representative example from the development set of RV segmentation on the LA and SA images, showing how Tempera is able to delineate the RV successfully.}
  \label{fig:segmentation_1}
\end{figure}

A representative segmentation of the SA and LA images can be seen in \textit{Figure \ref{fig:segmentation_1}}. Tempera's segmentations closely match the ground truth labels, except in the LA basal slices, where Tempera occasionally under-segments the RV.

\begin{table}[ht!]
\centering
\begin{tabularx}{\textwidth}{|Y|Y|Y|Y|Y|}
\hline
\multirow{2}{*}{\textbf{Dataset}} & \multicolumn{2}{Y|}{\textbf{Dice Score}} & \multicolumn{2}{l|}{\textbf{Hausdorff Distance (mm)}} \\ \cline{2-5} 
                                  & \textbf{SA}          & \textbf{LA}         & \textbf{SA}               & \textbf{LA}              \\ \hline
Validation                          & 0.895 $\pm$ n/a                & 0.829 $\pm$ n/a             & 11.8 $\pm$ n/a                     & 15.118 $\pm$ n/a                    \\ \hline
Testing                        & 0.836 $\pm$ 0.23                & 0.798 $\pm$ 0.28               & 26.31 $\pm$ 52.68                      & 31.19 $\pm$ 71.29                     \\ \hline
Testing (no failures)                       & 0.896 $\pm$ 0.05                 & 0.899 $\pm$ 0.06               & 18.384 $\pm$ 24.99                    & 14.186 $\pm$ 26.86                    \\ \hline
\end{tabularx}%
\caption{Summary of Tempera's performance on the (competition) validation and test sets using Dice score and Hausdorff distance. The test set's evaluation was repeated (no failures), in which we removed cases where the preprocessing step failed.}
\label{tab:results}
\vspace{-16mm}
\end{table}


\section{Discussion}
Tempera network can perform RV segmentation well on a challenging dataset. By designing a model that interchanges between 2D and 3D space, we can reap several benefits. Using the SA as 2D slices rather than just a 3D volume, we are able to effectively increase the number of available training data. On the other hand, when using 3D operations the network has the option to utilize structural information between slices to improve its performance. Finally, we exploit these geometric redundancies to improve the LA segmentations by using the GST.


The preprocessing ROI detection and standardization is an important step when dealing with data variation from different scan vendors. By detecting a small ROI which contains the RV, we can minimize erroneous segmentations, minimize the network learning irrelevant background features and allow for a smaller model. However, the detection may fail in the presence of pathology, artefacted or low SNR images and, in future work, we plan to improve this heart detection step. Short-term it can benefit from restricting the detection within a central region of the image. Additionally, we can take advantage of the spatial alignment of the LA and SA to project the centres from one image space to the other. In the long term, it can be improved by using a neural network-based detector.


There has been limited work done on incorporating the multiple views of cardiac MRI to improve the segmentation of cardiac chambers. \cite{Chen2019} \textit{et al} use the long axis views as shape priors for their short-axis cardiac segmentation. However, they restricted their investigations to the left myocardium, whose shape is much more regular than the RV's. Thus, their approach cannot be directly applied to the RV. Furthermore, the authors of \cite{uunet} focus on preprocessing and data augmentation techniques to solve the data domain shift presented between different scan vendors and clinics, as part of the first M\&Ms challenge \cite{9458279}. For the RV segmentation, they obtained an average Dice score of 0.88.

\section{Conclusions}
For the M\&Ms2 challenge, we propose the Tempera network to segment the right ventricle in the SA and LA cardiac MRI. Tempera uses both views and shares the learned features between them through hard-weight layer sharing in 2D space. A non-trainable spatial transformer, GST, allows for efficient sharing of information between the cardiac views. 

Our Tempera network can seamlessly be extended to also segment the LV and the myocardium. Additionally, it can be updated to incorporate other chamber views, such as 2-Chamber and 3-Chamber. The only modification needed for this is the inclusion of additional convolutional branches.

Furthermore, we can improve the robustness of the network by utilising the temporal information (ED and ES). We can compute the deformation matrix across the cardiac cycle. Then, we can pass the deformation matrix to the GST to associate the temporal information between the ED and ES phases.



\section*{Acknowledgements}
This work was supported by the UKRI CDT in AI for Healthcare http://ai4health.io (Grant No. EP/S023283/1) and the British Heart Foundation Centre of Research Excellence at Imperial College London (RE/18/4/34215).

%
%
%
\bibliographystyle{splncs04}
\bibliography{main.bib}

\end{document}